%% file: main.tex
\documentclass[acmsmall, screen, nonacm]{acmart}
\usepackage{subcaption}
\authorsaddresses{}

\AtBeginDocument{%
  \providecommand\BibTeX{{%
    \normalfont B\kern-0.5em{\scshape i\kern-0.25em b}\kern-0.8em\TeX}}}
\begin{document}

%%
%% The "title" command has an optional parameter,
%% allowing the author to define a "short title" to be used in page headers.
\title{Don't Look at the Camera: Achieving Perceived Eye Contact}

%%
%% The "author" command and its associated commands are used to define
%% the authors and their affiliations.
%% Of note is the shared affiliation of the first two authors, and the
%% "authornote" and "authornotemark" commands
%% used to denote shared contribution to the research.

\author{Alice Gao}
\affiliation{
  \institution{University of Washington}
  \city{Seattle}
  \country{atgao@cs.washington.edu}
}
\authornote{Denotes co-first authors.}

\author{Samyukta Jayakumar}\authornotemark[1]
\affiliation{
  \institution{University of California, Riverside}
  \city{Riverside}
  \country{sjaya012@ucr.edu}}
% \authornote{Denotes co-first authors.}

\author{Marcello Maniglia}\authornotemark[1]
\affiliation{
  \institution{University of California, Riverside}
  \city{Riverside}
  \country{mmanig@ucr.edu}
}
% \authornote{Denotes co-first authors.}

\author{Brian Curless}
\affiliation{
  \institution{University of Washington}
  \city{Seattle}
  \country{curless@cs.washington.edu}
}

\author{Ira Kemelmacher-Shlizerman}
\affiliation{
  \institution{University of Washington}
  \city{Seattle}
  \country{kemelmi@cs.washington.edu}}

\author{Aaron R. Seitz}
\affiliation{
  \institution{Northeastern University}
  \city{Boston}
  \country{a.seitz@northeastern.edu}}

\author{Steven M. Seitz}
\affiliation{
  \institution{University of Washington}
  \city{Seattle}
  \country{seitz@cs.washington.edu}
}

\renewcommand{\shortauthors}{Gao, Jayakumar, Maniglia, et al.}
% \clearpage
% \authornote{Denotes co-first authors.}
%%
%% By default, the full list of authors will be used in the page
%% headers. Often, this list is too long, and will overlap
%% other information printed in the page headers. This command allows
%% the author to define a more concise list
%% of authors' names for this purpose.
% \renewcommand{\shortauthors}{Trovato and Tobin, et al.}

%%
%% The abstract is a short summary of the work to be presented in the
%% article.
\begin{abstract}
We consider the question of how to best achieve the perception of eye contact when a person is captured by camera and then rendered on a 2D display. For single subjects photographed by a camera, conventional wisdom tells us that looking directly into the camera achieves eye contact. Through empirical user studies, we show that it is instead preferable to {\em look just below the camera lens}. We quantitatively assess where subjects should direct their gaze relative to a camera lens to optimize the perception that they are making eye contact.
\end{abstract}

%%
%% The code below is generated by the tool at http://dl.acm.org/ccs.cfm.
%% Please copy and paste the code instead of the example below.
%%
% \begin{CCSXML}
% <ccs2012>
%    <concept>
%        <concept_id>10003120.10003121.10011748</concept_id>
%        <concept_desc>Human-centered computing~Empirical studies in HCI</concept_desc>
%        <concept_significance>500</concept_significance>
%        </concept>
%  </ccs2012>
% \end{CCSXML}

% \ccsdesc[500]{Human-centered computing~Empirical studies in HCI}

\begin{teaserfigure}
\centerline{
  \includegraphics[width=\textwidth]{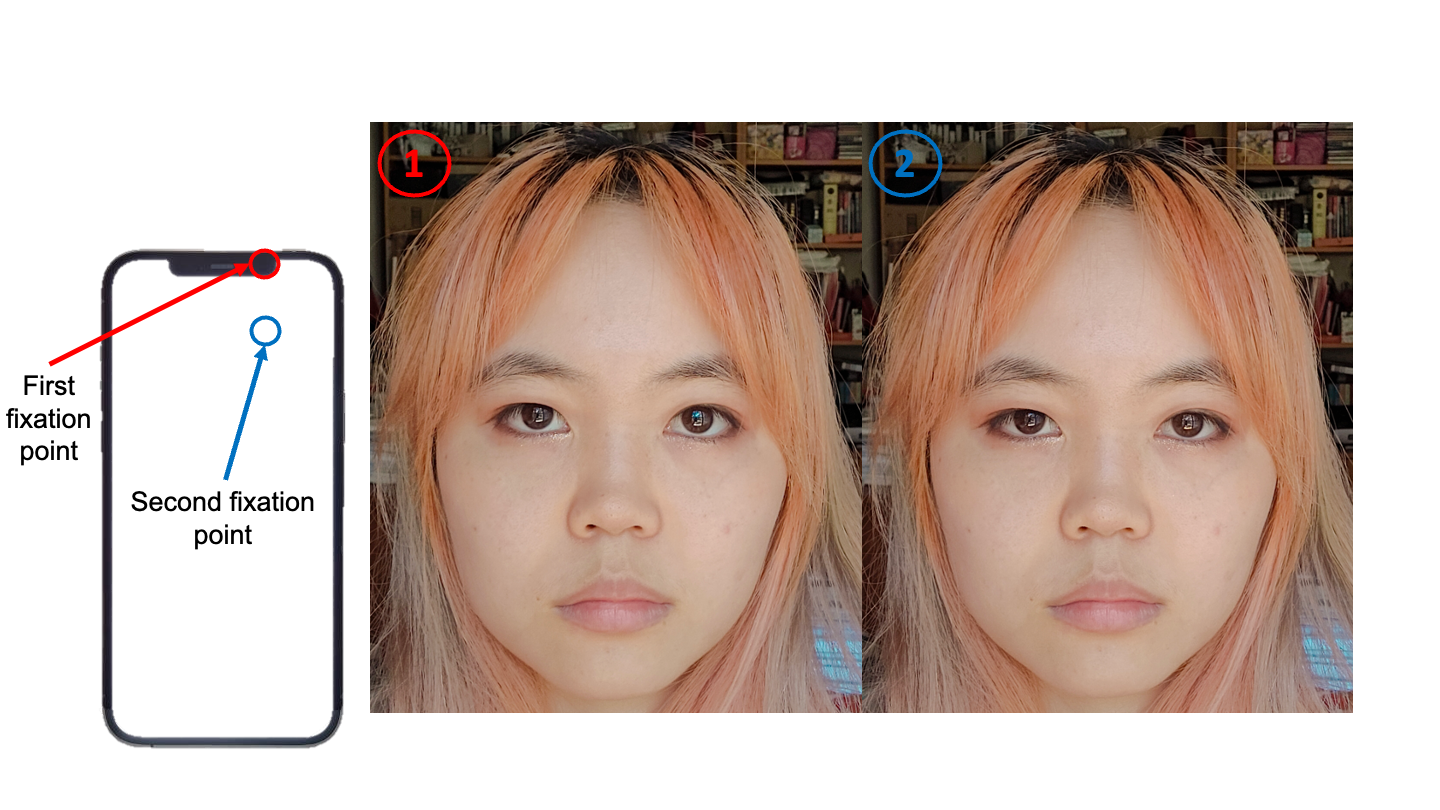}}
  \caption{(left) Directing our gaze at a camera lens is perceived as looking slightly upwards. (right) Looking two degrees of visual angle below the camera lens yields a truer perception of eye contact.}
  \Description{selfies}
  \label{fig:teaser}
\end{teaserfigure}

\maketitle

\input{files/1_intro}

\input{files/2_relatedworks}

\input{files/3_study}

\input{files/4_discussion}

\input{files/5_conclusion}

%%
%% The next two lines define the bibliography style to be used, and
%% the bibliography file.
\bibliographystyle{ACM-Reference-Format}
\bibliography{bibfile}

%%
%% If your work has an appendix, this is the place to put it.
% \input{files/appendix}

\end{document}

%% file: files/1_intro.tex
\section{Introduction}\label{sec:intro}
%sandy{The Intro, as written, lacked motivation and grounding. I found this under your Related Work first paragraph, which I moved here to better establish the context and significance of your work.}

Gaze is an essential social cue for person-person interaction, long-studied in vision science. Beyond indicating an individual's focus of attention, gaze is crucial for establishing joint attention~\cite{bayliss2011direct, moore2014joint, stephenson2021gaze}, exerting social influence (e.g., persuasion~\cite{kleinke1986gaze}), and deducing others’ mental states~\cite{baron1995children, calder2003disgust}. It is especially important in conversational settings since gaze cues can enhance verbal communication, including regulating turn-taking in conversation \cite{argyle1976gaze} and facilitating instructions \cite{andrist_bidirectional}. 

Today, people frequently interact through teleconferencing systems. Here, we ask the  question: where should a user look relative to the camera to give the impression of making eye contact?  The seemingly intuitive answer is to look into the camera's `eye,' or lens. Surprisingly, we show this to be incorrect: the optimal fixation point for the perception of eye contact is {\em just below the camera lens} (Fig. \ref{fig:teaser}).

%% file: files/2_relatedworks.tex
\section{Related Work}\label{sec:relatedworks}

\subsection{Gaze Awareness}
%\sandy{This is the par I moved to the Introduction since it serves as great motivation for your work.}
%Gaze is a long-studied subject in vision science due to its importance as a social cue for interaction. Beyond merely indicating an individual's focus of attention, gaze is crucial for establishing joint attention \cite{bayliss2011direct, moore2014joint, stephenson2021gaze}, exerting social control (such as during intimidation or persuasion \cite{kleinke1986gaze}), and deducing others’ mental state \cite{baron1995children, calder2003disgust}. It is especially important in conversational settings since gaze cues can enhance verbal communication, including the regulating turn-taking in conversation \cite{argyle1976gaze} and facilitating instructions \cite{andrist_bidirectional}. 

Many studies have addressed gaze perception and eye contact in face-to-face and video-mediated conversation. However, we note that the exact definition of eye contact is difficult to define due to its subjective nature. Various studies analyzing gaze perception have shown a large variation in definition of eye contact and the exact degree of involvement of the two parties needed for gaze perceptions \cite{Jongerius2020}. Gale and Monk \cite{gale-monk} differentiate three categories within gaze awareness to show that under different configurations of image size and camera placement, it is possible to perceive eye contact in video-mediated settings. The three categories are: {\em full gaze awareness}, which is the knowledge of what object in the environment someone is looking at; {\em partial gaze awareness}, the knowledge of the general direction someone is looking in; and {\em mutual gaze awareness}, the knowledge of whether someone is looking at you, i.e.,  what is typically thought of as eye contact.  

Previous studies have established that humans are excellent differentiators of gazes that are averted or directed somewhere on the face. Gibson and Pick \cite{gibson1963perception} measured the subjective perception of eye contact when the gazer fixated on points horizontally displaced from the subject's face. They observed that people are extremely sensitive to eye contact perception, which is comparable with visual acuity. %While these results hold true for fixations spaced at large visual angles, 
Further studies tested whether the same could be observed when the fixations were closely spaced \cite{cline1967perception, knight1973eye}. Kruger and Huckstedt conducted an experiment where the gazer fixated on seven points around the eyes of the perceiver (forehead, bridge and tip of nose, left and right eye, left and right edges of the face) and found that perceivers were able to identify the different fixations, albeit not with high accuracy \cite{kruger1969evaluation}. These paradigms were also explored in the realm of video conferencing platforms. Multiple studies examined perceived eye contact when gazers looked above or below the camera lens \cite{stapley1972visual, stokes1969human, white1970eye}. These studies collectively observed that the threshold of losing eye contact perception was 4.5 degrees above and 5.5 degrees below the camera. The studies reported that the sensitivity to eye contact is roughly symmetric in all directions of fixations \cite{anstis1969perception, knight1973eye, stapley1972visual, stokes1969human}. 

Additionally, Chen \cite{chen2002leveraging} conducted a study that controlled for the gazer and gazer’s eye appearance to understand the sensitivity to gaze perception in a video-mediated setting. This study found that: (1) when perceivers viewed a looped video of the gazer fixating on different points, they were less sensitive to fixations made below the camera compared to fixations above, left and right of the camera, and (2) during an online conversation between the gazer and perceivers, the perceivers were more likely to perceive  eye contact, even at larger degrees of fixation below the camera. 

%Taking these past studies into account, our study investigates the peak of the optimal perception of eye contact along the vertical axis.

Though prior work established the importance of eye contact and quantified sensitivities to variations in gaze, they did not answer the question of where specifically to direct one's gaze relative to a camera to optimize eye contact, an important case in practice. Further, past studies neither employed measures of confidence about the gaze (exactly where the gazer is looking during image/video acquisition) nor controlled for the gazer's head movements, which was previously shown to influence eye contact perception \cite{anstis1969perception, cline1967perception, gibson1963perception}. 

Here we present a carefully-designed user study to measure camera-relative gaze and perceived eye contact. Specifically, we took images of  actors looking at fixation points above, at, and below a camera placed at typical monitor distance (20”-24”), and asked human subjects to indicate which image(s) correspond to the actor making eye contact.  We find that the images that led to the largest perception of eye contact were not those in which the actors were directly looking at the camera, as one might expect, but actually on the bridge of the nose about two degrees of visual angle (dva or °) below the center of the lens of the camera.

%% file: files/3_study.tex
\section{Gaze Perception} \label{sec:gaze}
% Our goal is to carefully quantify the perception of gaze and answer the question: where should a person look relative to the camera to be perceived as making eye contact? 

\subsection{Experiment Design}
%Taking into account these differences, our work focuses on building a system and presenting a novel and generalizable display configuration of participants' videos to convey gaze cues and enable a gaze aware video conferencing system.
We designed an experiment to quantify the experience of eye contact along the vertical direction with a gaze perception paradigm that:
\begin{enumerate}
    \item Acquires fixations that are a minimum of 1 degree of visual angle apart,
    \item Controls for any extraneous eye or head movements during image acquisition from gazers, and
    \item Uses a self-paced design to acquire a perceiver's subjective measure of eye contact.
\end{enumerate}  

\subsection{Experiment Setup}
\begin{figure*}[!h]
  \centering
  \includegraphics[width=0.8\linewidth]{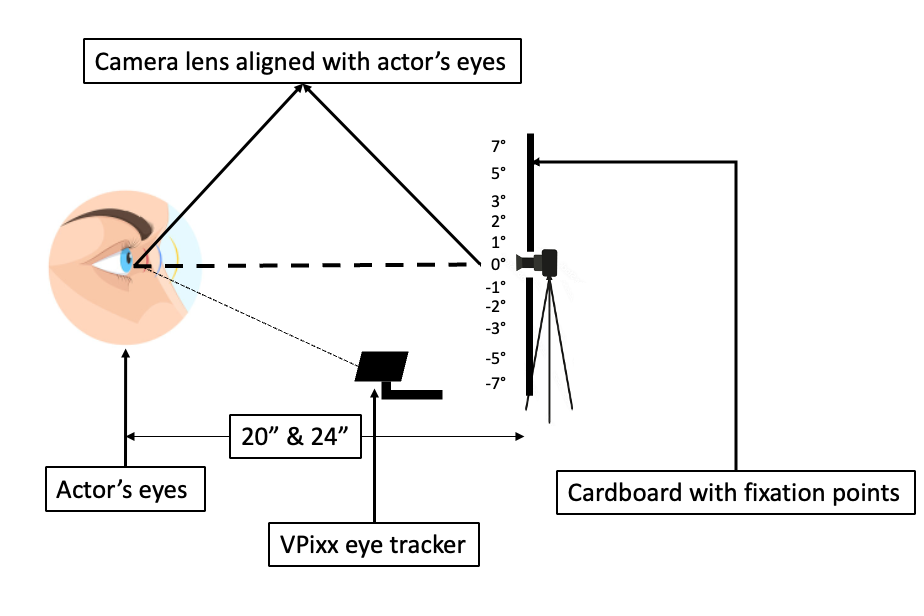}
  \caption{Equipment set up used to collect pictures from gazers.}
  \Description{Gaze study setup }
  \label{fig: gaze study schematic}
\end{figure*}
\subsubsection{Participants}
Seventeen participants (mean age: 22.3 ± SD 5.2 years; 7 females) with normal or corrected-to-normal vision were recruited to take part in the study. Experiment protocols were approved by UCR Institutional Review Board, and all participants gave written informed consent prior to the experiment.

\subsubsection{Stimuli}
Stimuli consisted of a total of 88 pictures of gazers' faces that were collected using a Logitech C920x HD Pro Webcam. We refer to gazers as \textit{actors} in our experiment and for the remainder of the paper. These pictures were collected from four different actors 
(3 male, 1 female, mean age = 25+/-2.3) using an eye-tracking system that let us verify the actors were correctly gazing at intended locations on the screen. 

The actors were instructed to fixate their gaze on a series of fixation dots on a custom-made cardboard structure, as shown in Fig. \ref{fig: gaze study schematic}. The camera was aligned with the actor's eye level, and pictures were collected while the actor fixated on each of 11 fixation locations (-7°, -5°, -3°, -2°, -1°, 0°, 1°, 2°, 3°, 5°, 7°) vertically displaced from central fixation (0° - camera lens).  

Prior to picture collection, a 9-point calibration and validation procedure was performed. Cutoff accuracy to accept the calibration was 0.5°. Calibration and eye monitoring was performed using a Vpixx Trackpixx system \cite{vpixx} set up on a Display++ monitor \cite{Display++}. After calibration, the monitor was  replaced by a custom-made cardboard structure (Fig. \ref{fig: gaze study schematic}) with an aperture at the center within which the webcam was placed. Haircross stickers were placed at the 11 fixation locations mentioned above. 

Actors were then asked to fixate at different fixation locations, and a 3-second countdown was provided to inform them about the exact moment the picture would be taken. We used online eye tracking with the Vpixx system to ensure that the actor was looking at the intended spot. If the actor’s eyes moved >0.5° of the intended spot at the moment of picture collection, the image was discarded and reacquired until fixation was within 0.5°. This procedure was repeated for two viewing distances (20” and 24” \footnote[1]{The \href{https://workerscomp.nm.gov/sites/default/files/documents/publications/Ergonomics/Info_Ergonomics_Computer_Monitor_Positioning.pdf}{general guideline} is to have the monitor 20 to 40 inches away from the eyes.}), and the fixation points were adjusted accordingly to maintain consistent angular spacing. 

Viewing distances were selected based on typical distances at which desktops and laptops would be placed during video calls. External lighting was adjusted to ensure equal luminance and exposure across actors. The images were then manually cropped and resized so that the actors' eyes in all pictures were aligned at the center of the screen for the main the experiment. 

\subsection{Methodology}
\begin{figure*}[!h]
  \centering
  \includegraphics[width=\linewidth]{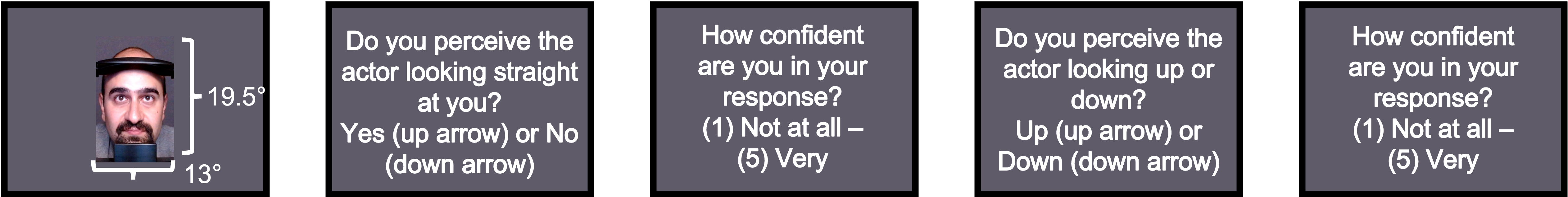}
  \caption{Single trial during data collection for reporting subjective gaze perception. All trials were self-paced, and participants answered all questions for each picture they viewed. %\brian{say the left box is an image presented to the user and the three other boxes are questions in the study.  If space/length is an issue, could combine this fig with the other, just adding the image as (b) in the figure and stating that it is a captured image as shown to the user in a study, omit the questions which already appear in the text.}
  }
  \Description{Gaze study setup}
  \label{fig: gaze qs}
\end{figure*}

\subsubsection{Data Collection}
Images were randomized and presented on a Samsung S24e310 LCD monitor with a resolution of 1920x1080. Participants sat in a dark room, 23’’ (60 cm) from the screen. This distance was chosen to keep the size of the image the same as the face of the actor in the images shown. A total of 176 trials were presented, where participants viewed each of the 88 pictures twice and were asked to answer two questions per picture (Fig. \ref{fig: gaze qs}). The exact phrasing of the questions was as follows (Fig. \ref{fig: gaze qs}): \textbf{‘Do you perceive the actor looking straight at you?’} and \textbf {‘Do you perceive the actor looking up or down?’}. 

Participants responded to the questions using the arrow keys on the keyboard, i.e., up arrow (for ‘Yes’ to the first question and ‘Up’ for the second question) or down arrow (for ‘No’ to the first question and ‘Down’ for the second question). Additionally, they were also asked to provide a confidence rating on a 5-point scale (1 - least confident and 5 - very confident) for their responses to each question. Picture size was 19.5° vertical x 13° horizontal. Visual stimuli were generated using the Psychophysics Toolbox for Matlab \cite{brainard1997psychophysics, pelli1997videotoolbox}.
\subsubsection{Data Analysis}
Data was collapsed across the two viewing distances and analyzed separately for the ‘yes vs no’ and the ‘up vs down’ questions. In the ‘yes vs no’ case, ‘yes’ responses per eye offset were fitted to a Gaussian distribution for each participant, and individual peaks of the function were compared against zero with a one sample t-test. In the ‘up vs down’ case, a psychometric function was fitted to the data corresponding to the ‘down’ response across the various eye offsets. Gaussian functions were fit to individual distributions of ‘yes’ and ‘down’ responses across fixation offsets. The peak of the function was estimated and used as the individual subjective gaze offset that led to the perception of eye contact.

%\brian{Are the 20" and 24" results just lumped together?}

\begin{figure*}[h]
  \centering
  \includegraphics[width = \linewidth]{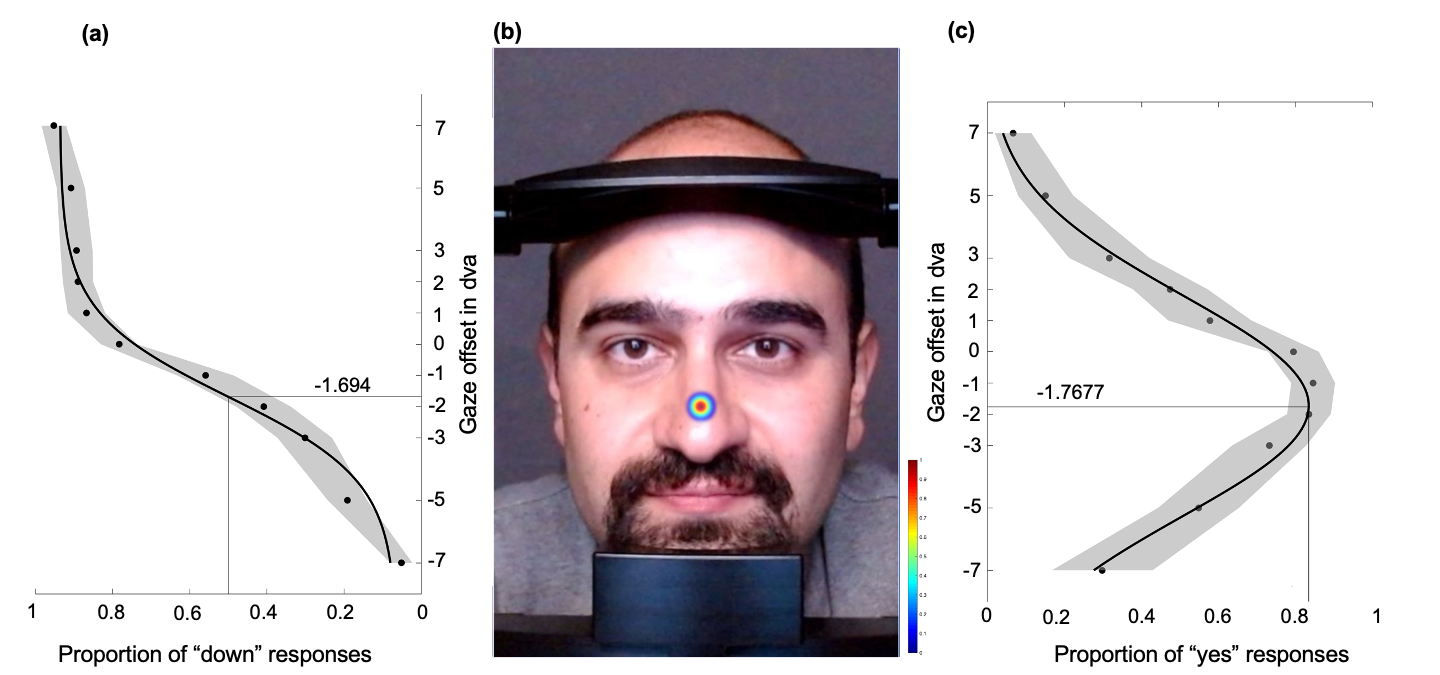}
  \caption{(a) Average distribution of ‘yes’ response across gaze offset with 95\% CI and Gaussian fit showing a peak at 1.76° below 0° fixation (negative numbers indicate fixations above the center of the camera (0°), and positive numbers indicate fixations below the center). (b) Visual representation of the point of fixation (-1.76°) on an actor’s face to produce an impression of eye contact. The circular heatmap on the image of the actor corresponds to the Gaussian fit of perceived eye contact. The red center of the heatmap is the optimal perceived point of eye contact. As one moves further away from this point, towards the outer blue ring, so does the perception of eye contact being made. (c) Average distribution of ‘down’ response across gaze offset fitted with a psychometric function. ‘dva' stands for degrees of visual angle. }
  \label{fig: gaze-results}
\end{figure*}

\subsection{Results}

Results are shown in Fig. \ref{fig: gaze-results}, separating the yes-no and the up-down study questions. 

\textbf{Yes-no.} The shift in perceived gaze from zero (central fixation) was estimated by fitting a Gaussian distribution to the proportion of ‘yes’ responses (i.e., ‘the actor is looking at me’) for the various gaze offset levels (Fig. \ref{fig: gaze-results}a). The offset corresponding to the peak of the distribution was estimated and tested against zero via a one sample t-test. Results showed a significant shift from zero (t(16) = 4.45, p < 0.0001).

\textbf{Up-down.} The shift in perceived gaze from zero (central fixation) was estimated by fitting a psychometric function to the proportion of responses ‘down’ for the various gaze offset levels (Fig. \ref{fig: gaze-results}c). The point of subjective equality (or PSE, where participants responded 50\% up and 50\% down for a specific gaze offset) was estimated and tested against zero via a one sample t-test. Results showed a significant shift from zero (t(16) = 6.19, p < 0.0001).

% \textbf{

%Our social interactions rely upon our ability to recognize a variety of facial cues that provide information required to understand our conversation partner's identity and emotion, not just our ability to perceive eye contact. These social interactions are best perceived when gazing halfway down the nose \cite{peterson_eckstein}. When one foveates just below the eyes, they are able to perceive cues from another's eyes, nose, and mouth, enabling them to maximize the number of facial cues they perceive. Thus, from an evolutionary perspective, it is likely that we have learned to categorize gazes below eye-level as being eye-contact. We also note that this is consistent with other areas of human perception, such as phonetic discrimination that underlies our ability to understand language \cite{Kuhl1992-ld}. 

%% file: files/4_discussion.tex
\section{Discussion}\label{sec:discussion}

In this study, we aimed to estimate the vertical gaze offset that leads to the best experience of eye contact in conditions of simulated remote video calls. To do so, we collected the subjective gaze perception of participants while they viewed images of actors fixating on vertically displaced points above and below the center camera. Results from our study point to a consistent trend across all participants perceiving the actor as making eye contact when the actor fixated at a point that was approximately 2° below the center fixation point (camera) when seated 23" away from the screen. We also observed a relatively high proportion of ‘yes’ responses between 0° and 3° fixation points (camera and below the camera). These results are represented by the heatmap in Fig. \ref{fig: gaze-results}b): the red center of the heatmap is the peak of the Gaussian distribution and the peak of perceived eye contact. As the actor's gaze moves away from this point, so does the perceived sensation of eye contact. Overall, participants tended to perceive the actor's downwards gaze as eye contact. Our results indicate that the perception of eye contact is complex and non-intuitive.  We assume that eye contact is achieved between people when looking into the ‘camera’ of their eyes, and thus looking directly into an actual camera would intuitively result in an image conveying eye contact.  However, our results here show that this is not the case, and, for a camera placed at eye-level, we should look {\em below}, not into the camera lens, to optimize the perception of eye contact. 

These results are consistent with prior studies that showed a range of tolerance for eye offset in an actor to elicit perception of eye contact when the actor was looking into the camera or points below the camera (up to approximately 7°) with above chance levels. However, our study challenges this viewpoint by observing this effect only for smaller windows of fixations (up to approximately 3° above chance). This could be attributed to differences in the experiment paradigm, as follows:

\begin{enumerate}
    \item Prior studies did not employ measures to ensure the actor was fixating on the intended point before collecting the image or video. Our study employs eye tracking to ensure that the eyes were fixating the intended spot at the time of the picture being taken. .
    \item Our study additionally uses head and chin rests to avoid head movements that could affect perceived eye contact.
    \item Prior studies, for both in-person and video conference settings, did not control for whether the eye level of the actor and perceiver were aligned. This could have contributed to the larger bias in eye contact perception. Our study controls for this by first aligning the actor's eye with the center of the camera during image acquisition and by aligning the perceiver's eyes with the eyes of the actors image. 
\end{enumerate}
% \atgaocomment{make sure to double check the snap to contact bit later, reviewer note about some confusion here}

Past studies attributed this high level of subjective perception of eye contact when the actor looks below the camera to the eye's anatomy \cite{stapley1972visual}. The position of the iris within the sclera changes noticeably when the actor looks above the camera because the upper eyelids follow that change while the lower eyelids stay fixed; the position of the iris in relation to the sclera does not change noticeably when the actor looks below the camera. This could also be explained by the Snap to Contact Theory, proposed by Chen, which states that unless perceivers are absolutely certain that the actors are not looking at them, they will bias their perception towards eye contact \cite{chen2002leveraging}. That is, due to the less noticeable shift in the appearance of the sclera when a person looks downwards, observers will tend to  perceive eye contact. However, a complementary explanation better matches the cone of equivalent gaze, as seen for horizontal eye-movements \cite{cone_of_gaze}. Gamer and Hecht originally introduced the metaphor of  cone of gaze to represent the range of gaze directions made by a gazer that are classified as the perciever as engaging in mutual gaze, or making eye contact.

Taken together with prior work, our results suggest that gaze that is slightly downwards from the eyes and toward the middle of the nose is perceived in face-to-face interactions as making eye contact.  Why might this be?  Our social interactions rely upon our ability to recognize a variety of facial cues that provide information required to understand our conversation partner's identity and emotion, not just our ability to perceive eye contact. These social interactions are best perceived when gazing halfway down the nose \cite{peterson_eckstein}. When one foveates just below the eyes, they are able to perceive cues from another's eyes, nose, and mouth, enabling them to maximize the number of facial cues they perceive. Thus, from an evolutionary perspective, it is likely that we have learned to categorize gazes below eye-level as being eye-contact. We also note that this is consistent with other areas of human perception, such as phonetic discrimination that underlies our ability to understand language \cite{Kuhl1992-ld}.

\subsection{Applications}
Our result that perceived eye contact through a camera occurs when a user looks approximately 2\textdegree \space below the camera, at least for 20"-24" distances from the camera, has a variety of useful applications. First, we can facilitate photographs with better perceived eye contact by instructing actors to gaze at the optimal point. This includes selfies, which are typically shot at an arm's length away, similar to the distance between camera and actor in our study. Second, we can improve gaze cues in video conferencing systems. Leveraging the fact that the point of best perceived eye contact is slightly below the webcam, future gaze correction systems could redirect participants' gazes correspondingly in the vertical axis, allowing for the optimal perception of eye contact. 
%\steve{removed subsequent text which seemed to repeat this point or go into details that I think are less pertinent to the result}
%It is also possible to create {\em gaze aware} video conferencing systems with this result. Using computer vision techniques one can synthesize the video of a user from a fixed location two degrees above the optimal location of perceived eye contact, for instance, in between the eyes of a participant's video feed. This way, when a user does engage in eye contact with a participant and looks halfway down there nose, there is optimal perceived eye contact. Additionally, this also captures natural eye saccades and gaze cues as it also correctly captures the user looking away.

\subsection{Limitations and Future Work}
Though our experiment quantified the degree of visual angle for best perceived eye contact, or mutual gaze, and verified this through the use of an eye tracker, some limitations remain. Our study did not account for different variables, such as the size of monitors, seating distances from the monitor, or the location of the webcam relative to the actor's eyes and the impact of these variables on perception of eye contact. Future work could investigate  the effects of these variations. 

Additionally, our experiment focused on still images. These do not capture the entire possible set of gazes during actual conversation in a video-mediated setting. Nor do they capture which set of gazes would be perceived as eye contact in this scenario. Further, we do not quantify how long participants {\em want} to engage in mutual gaze throughout a conversation. A potential future study could engage two actors in a short, scripted conversation and use an eye tracker to verify their gaze. Then, after the conversation, researchers could ask them at which duration they felt like the other actor was making eye contact. This would help us better understand how people perceive gaze cues throughout video interaction.

%% file: files/5_conclusion.tex
%\subsection{Conclusion}
%We show that the best perceived eye contact through a camera arises when a user looks approximately 2\textdegree \space below the camera. %\steve{only applies to a specific screen distance, right?}%. %This result is useful for a variety of camera- and video-mediated settings. We believe our result will enable video conferencing system developers to incorporate a gaze correction component into their systems that will allow for natural and intuitive gaze cues as well as the perception of eye contact. 

%% file: main.bbl
%%% -*-BibTeX-*-
%%% Do NOT edit. File created by BibTeX with style
%%% ACM-Reference-Format-Journals [18-Jan-2012].

\begin{thebibliography}{26}

%%% ====================================================================
%%% NOTE TO THE USER: you can override these defaults by providing
%%% customized versions of any of these macros before the \bibliography
%%% command.  Each of them MUST provide its own final punctuation,
%%% except for \shownote{}, \showDOI{}, and \showURL{}.  The latter two
%%% do not use final punctuation, in order to avoid confusing it with
%%% the Web address.
%%%
%%% To suppress output of a particular field, define its macro to expand
%%% to an empty string, or better, \unskip, like this:
%%%
%%% \newcommand{\showDOI}[1]{\unskip}   % LaTeX syntax
%%%
%%% \def \showDOI #1{\unskip}           % plain TeX syntax
%%%
%%% ====================================================================

\ifx \showCODEN    \undefined \def \showCODEN     #1{\unskip}     \fi
\ifx \showDOI      \undefined \def \showDOI       #1{#1}\fi
\ifx \showISBNx    \undefined \def \showISBNx     #1{\unskip}     \fi
\ifx \showISBNxiii \undefined \def \showISBNxiii  #1{\unskip}     \fi
\ifx \showISSN     \undefined \def \showISSN      #1{\unskip}     \fi
\ifx \showLCCN     \undefined \def \showLCCN      #1{\unskip}     \fi
\ifx \shownote     \undefined \def \shownote      #1{#1}          \fi
\ifx \showarticletitle \undefined \def \showarticletitle #1{#1}   \fi
\ifx \showURL      \undefined \def \showURL       {\relax}        \fi
% The following commands are used for tagged output and should be
% invisible to TeX
\providecommand\bibfield[2]{#2}
\providecommand\bibinfo[2]{#2}
\providecommand\natexlab[1]{#1}
\providecommand\showeprint[2][]{arXiv:#2}

\bibitem[Andrist et~al\mbox{.}(2017)]%
        {andrist_bidirectional}
\bibfield{author}{\bibinfo{person}{Sean Andrist}, \bibinfo{person}{Michael
  Gleicher}, {and} \bibinfo{person}{Bilge Mutlu}.}
  \bibinfo{year}{2017}\natexlab{}.
\newblock \showarticletitle{Looking Coordinated: Bidirectional Gaze Mechanisms
  for Collaborative Interaction with Virtual Characters}. In
  \bibinfo{booktitle}{\emph{Proceedings of the 2017 CHI Conference on Human
  Factors in Computing Systems}} (Denver, Colorado, USA)
  \emph{(\bibinfo{series}{CHI '17})}. \bibinfo{publisher}{Association for
  Computing Machinery}, \bibinfo{address}{New York, NY, USA},
  \bibinfo{pages}{2571–2582}.
\newblock
\showISBNx{9781450346559}
\urldef\tempurl%
\url{https://doi.org/10.1145/3025453.3026033}
\showDOI{\tempurl}


\bibitem[Anstis et~al\mbox{.}(1969)]%
        {anstis1969perception}
\bibfield{author}{\bibinfo{person}{Stuart~M Anstis}, \bibinfo{person}{John~W
  Mayhew}, {and} \bibinfo{person}{Tania Morley}.}
  \bibinfo{year}{1969}\natexlab{}.
\newblock \showarticletitle{The perception of where a face or
  television'portrait'is looking}.
\newblock \bibinfo{journal}{\emph{The American journal of psychology}}
  \bibinfo{volume}{82}, \bibinfo{number}{4} (\bibinfo{year}{1969}),
  \bibinfo{pages}{474--489}.
\newblock


\bibitem[Argyle and Cook(1976)]%
        {argyle1976gaze}
\bibfield{author}{\bibinfo{person}{Michael Argyle} {and} \bibinfo{person}{Mark
  Cook}.} \bibinfo{year}{1976}\natexlab{}.
\newblock \showarticletitle{Gaze and mutual gaze.}
\newblock  (\bibinfo{year}{1976}).
\newblock


\bibitem[Baron-Cohen et~al\mbox{.}(1995)]%
        {baron1995children}
\bibfield{author}{\bibinfo{person}{Simon Baron-Cohen}, \bibinfo{person}{Ruth
  Campbell}, \bibinfo{person}{Annette Karmiloff-Smith}, \bibinfo{person}{Julia
  Grant}, {and} \bibinfo{person}{Jane Walker}.}
  \bibinfo{year}{1995}\natexlab{}.
\newblock \showarticletitle{Are children with autism blind to the mentalistic
  significance of the eyes?}
\newblock \bibinfo{journal}{\emph{British Journal of Developmental Psychology}}
  \bibinfo{volume}{13}, \bibinfo{number}{4} (\bibinfo{year}{1995}),
  \bibinfo{pages}{379--398}.
\newblock


\bibitem[Bayliss et~al\mbox{.}(2011)]%
        {bayliss2011direct}
\bibfield{author}{\bibinfo{person}{Andrew~P Bayliss}, \bibinfo{person}{Jessica
  Bartlett}, \bibinfo{person}{Claire~K Naughtin}, {and} \bibinfo{person}{Ada
  Kritikos}.} \bibinfo{year}{2011}\natexlab{}.
\newblock \showarticletitle{A direct link between gaze perception and social
  attention.}
\newblock \bibinfo{journal}{\emph{Journal of Experimental Psychology: Human
  Perception and Performance}} \bibinfo{volume}{37}, \bibinfo{number}{3}
  (\bibinfo{year}{2011}), \bibinfo{pages}{634}.
\newblock


\bibitem[Brainard and Vision(1997)]%
        {brainard1997psychophysics}
\bibfield{author}{\bibinfo{person}{David~H Brainard} {and}
  \bibinfo{person}{Spatial Vision}.} \bibinfo{year}{1997}\natexlab{}.
\newblock \showarticletitle{The psychophysics toolbox}.
\newblock \bibinfo{journal}{\emph{Spatial vision}} \bibinfo{volume}{10},
  \bibinfo{number}{4} (\bibinfo{year}{1997}), \bibinfo{pages}{433--436}.
\newblock


\bibitem[Calder(2003)]%
        {calder2003disgust}
\bibfield{author}{\bibinfo{person}{Andrew~J Calder}.}
  \bibinfo{year}{2003}\natexlab{}.
\newblock \showarticletitle{Disgust discussed.}
\newblock \bibinfo{journal}{\emph{Annals of Neurology}} (\bibinfo{year}{2003}).
\newblock


\bibitem[Chen(2002)]%
        {chen2002leveraging}
\bibfield{author}{\bibinfo{person}{Milton Chen}.}
  \bibinfo{year}{2002}\natexlab{}.
\newblock \showarticletitle{Leveraging the Asymmetric Sensitivity of Eye
  Contact for Videoconference}. In \bibinfo{booktitle}{\emph{Proceedings of the
  SIGCHI Conference on Human Factors in Computing Systems}} (Minneapolis,
  Minnesota, USA) \emph{(\bibinfo{series}{CHI '02})}.
  \bibinfo{publisher}{Association for Computing Machinery},
  \bibinfo{address}{New York, NY, USA}, \bibinfo{pages}{49–56}.
\newblock
\showISBNx{1581134533}
\urldef\tempurl%
\url{https://doi.org/10.1145/503376.503386}
\showDOI{\tempurl}


\bibitem[Cline(1967)]%
        {cline1967perception}
\bibfield{author}{\bibinfo{person}{Marvin~G Cline}.}
  \bibinfo{year}{1967}\natexlab{}.
\newblock \showarticletitle{The perception of where a person is looking}.
\newblock \bibinfo{journal}{\emph{The American journal of psychology}}
  \bibinfo{volume}{80}, \bibinfo{number}{1} (\bibinfo{year}{1967}),
  \bibinfo{pages}{41--50}.
\newblock


\bibitem[Gale and Monk(2012)]%
        {gale-monk}
\bibfield{author}{\bibinfo{person}{Caroline Gale} {and} \bibinfo{person}{Andrew
  Monk}.} \bibinfo{year}{2012}\natexlab{}.
\newblock \showarticletitle{Where am I looking? The accuracy of video-mediated
  gaze awareness}.
\newblock \bibinfo{journal}{\emph{Attention Perception \& Psychophysics}}
  \bibinfo{volume}{62} (\bibinfo{date}{04} \bibinfo{year}{2012}),
  \bibinfo{pages}{586--595}.
\newblock
\urldef\tempurl%
\url{https://doi.org/10.3758/BF03212110}
\showDOI{\tempurl}


\bibitem[Gamer and Hecht(2007)]%
        {cone_of_gaze}
\bibfield{author}{\bibinfo{person}{Matthias Gamer} {and} \bibinfo{person}{Heiko
  Hecht}.} \bibinfo{year}{2007}\natexlab{}.
\newblock \showarticletitle{Are you looking at me? Measuring the cone of gaze.}
\newblock \bibinfo{journal}{\emph{Journal of Experimental Psychology: Human
  Perception and Performance}} \bibinfo{volume}{33}, \bibinfo{number}{3}
  (\bibinfo{year}{2007}), \bibinfo{pages}{705–715}.
\newblock
\showISSN{1939-1277, 0096-1523}
\urldef\tempurl%
\url{https://doi.org/10.1037/0096-1523.33.3.705}
\showDOI{\tempurl}


\bibitem[Gibson and Pick(1963)]%
        {gibson1963perception}
\bibfield{author}{\bibinfo{person}{James~J Gibson} {and}
  \bibinfo{person}{Anne~D Pick}.} \bibinfo{year}{1963}\natexlab{}.
\newblock \showarticletitle{Perception of another person's looking behavior}.
\newblock \bibinfo{journal}{\emph{The American journal of psychology}}
  \bibinfo{volume}{76}, \bibinfo{number}{3} (\bibinfo{year}{1963}),
  \bibinfo{pages}{386--394}.
\newblock


\bibitem[Inc.(2022)]%
        {vpixx}
\bibfield{author}{\bibinfo{person}{VPixx~Technologies Inc.}}
  \bibinfo{year}{2022}\natexlab{}.
\newblock \bibinfo{title}{Vpixx}.
\newblock \bibinfo{howpublished}{\url{https://https://vpixx.com/}}.
\newblock


\bibitem[Jongerius et~al\mbox{.}(2020)]%
        {Jongerius2020}
\bibfield{author}{\bibinfo{person}{Chiara Jongerius}, \bibinfo{person}{Roy~S.
  Hessels}, \bibinfo{person}{Johannes~A. Romijn}, \bibinfo{person}{Ellen M.~A.
  Smets}, {and} \bibinfo{person}{Marij~A. Hillen}.}
  \bibinfo{year}{2020}\natexlab{}.
\newblock \showarticletitle{The Measurement of Eye Contact in Human
  Interactions: A Scoping Review}.
\newblock \bibinfo{journal}{\emph{Journal of Nonverbal Behavior}}
  \bibinfo{volume}{44}, \bibinfo{number}{3} (\bibinfo{date}{01 Sep}
  \bibinfo{year}{2020}), \bibinfo{pages}{363--389}.
\newblock
\showISSN{1573-3653}
\urldef\tempurl%
\url{https://doi.org/10.1007/s10919-020-00333-3}
\showDOI{\tempurl}


\bibitem[Kleinke(1986)]%
        {kleinke1986gaze}
\bibfield{author}{\bibinfo{person}{Chris~L Kleinke}.}
  \bibinfo{year}{1986}\natexlab{}.
\newblock \showarticletitle{Gaze and eye contact: a research review.}
\newblock \bibinfo{journal}{\emph{Psychological bulletin}}
  \bibinfo{volume}{100}, \bibinfo{number}{1} (\bibinfo{year}{1986}),
  \bibinfo{pages}{78}.
\newblock


\bibitem[Knight et~al\mbox{.}(1973)]%
        {knight1973eye}
\bibfield{author}{\bibinfo{person}{David~J Knight}, \bibinfo{person}{Daniel
  Langmeyer}, {and} \bibinfo{person}{David~C Lundgren}.}
  \bibinfo{year}{1973}\natexlab{}.
\newblock \showarticletitle{Eye-contact, distance, and affiliation: The role of
  observer bias}.
\newblock \bibinfo{journal}{\emph{Sociometry}} (\bibinfo{year}{1973}),
  \bibinfo{pages}{390--401}.
\newblock


\bibitem[Kr{\"u}ger and H{\"u}ckstedt(1969)]%
        {kruger1969evaluation}
\bibfield{author}{\bibinfo{person}{K Kr{\"u}ger} {and} \bibinfo{person}{B
  H{\"u}ckstedt}.} \bibinfo{year}{1969}\natexlab{}.
\newblock \showarticletitle{Evaluation of the direction of gazing}.
\newblock \bibinfo{journal}{\emph{Zeitschrift fur Experimentelle und Angewandte
  Psychologie}} \bibinfo{volume}{16}, \bibinfo{number}{3}
  (\bibinfo{year}{1969}), \bibinfo{pages}{452--472}.
\newblock


\bibitem[Kuhl et~al\mbox{.}(1992)]%
        {Kuhl1992-ld}
\bibfield{author}{\bibinfo{person}{P~K Kuhl}, \bibinfo{person}{K~A Williams},
  \bibinfo{person}{F Lacerda}, \bibinfo{person}{K~N Stevens}, {and}
  \bibinfo{person}{B Lindblom}.} \bibinfo{year}{1992}\natexlab{}.
\newblock \showarticletitle{Linguistic experience alters phonetic perception in
  infants by 6 months of age}.
\newblock \bibinfo{journal}{\emph{Science}} \bibinfo{volume}{255},
  \bibinfo{number}{5044} (\bibinfo{date}{Jan.} \bibinfo{year}{1992}),
  \bibinfo{pages}{606--608}.
\newblock


\bibitem[Moore et~al\mbox{.}(2014)]%
        {moore2014joint}
\bibfield{author}{\bibinfo{person}{Chris Moore}, \bibinfo{person}{Philip~J
  Dunham}, {and} \bibinfo{person}{Phil Dunham}.}
  \bibinfo{year}{2014}\natexlab{}.
\newblock \bibinfo{booktitle}{\emph{Joint attention: Its origins and role in
  development}}.
\newblock \bibinfo{publisher}{Psychology Press}.
\newblock


\bibitem[Pelli and Vision(1997)]%
        {pelli1997videotoolbox}
\bibfield{author}{\bibinfo{person}{Denis~G Pelli} {and}
  \bibinfo{person}{Spatial Vision}.} \bibinfo{year}{1997}\natexlab{}.
\newblock \showarticletitle{The VideoToolbox software for visual psychophysics:
  Transforming numbers into movies}.
\newblock \bibinfo{journal}{\emph{Spatial vision}}  \bibinfo{volume}{10}
  (\bibinfo{year}{1997}), \bibinfo{pages}{437--442}.
\newblock


\bibitem[Peterson and Eckstein(2012)]%
        {peterson_eckstein}
\bibfield{author}{\bibinfo{person}{Matthew~F. Peterson} {and}
  \bibinfo{person}{Miguel~P. Eckstein}.} \bibinfo{year}{2012}\natexlab{}.
\newblock \showarticletitle{Looking just below the eyes is optimal across face
  recognition tasks}.
\newblock \bibinfo{journal}{\emph{Proceedings of the National Academy of
  Sciences}} \bibinfo{volume}{109}, \bibinfo{number}{48}
  (\bibinfo{year}{2012}), \bibinfo{pages}{E3314--E3323}.
\newblock
\urldef\tempurl%
\url{https://doi.org/10.1073/pnas.1214269109}
\showDOI{\tempurl}


\bibitem[Stapley(1972)]%
        {stapley1972visual}
\bibfield{author}{\bibinfo{person}{Barry Stapley}.}
  \bibinfo{year}{1972}\natexlab{}.
\newblock \showarticletitle{Visual enhancement of telephone conversations}.
\newblock  (\bibinfo{year}{1972}).
\newblock


\bibitem[Stephenson et~al\mbox{.}(2021)]%
        {stephenson2021gaze}
\bibfield{author}{\bibinfo{person}{Lisa~J Stephenson},
  \bibinfo{person}{S~Gareth Edwards}, {and} \bibinfo{person}{Andrew~P
  Bayliss}.} \bibinfo{year}{2021}\natexlab{}.
\newblock \showarticletitle{From gaze perception to social cognition: The
  shared-attention system}.
\newblock \bibinfo{journal}{\emph{Perspectives on Psychological Science}}
  \bibinfo{volume}{16}, \bibinfo{number}{3} (\bibinfo{year}{2021}),
  \bibinfo{pages}{553--576}.
\newblock


\bibitem[Stokes(1969)]%
        {stokes1969human}
\bibfield{author}{\bibinfo{person}{R Stokes}.} \bibinfo{year}{1969}\natexlab{}.
\newblock \showarticletitle{Human factors and appearance design considerations
  of the mod II PICTUREPHONE{\textregistered} station set}.
\newblock \bibinfo{journal}{\emph{IEEE Transactions on Communication
  Technology}} \bibinfo{volume}{17}, \bibinfo{number}{2}
  (\bibinfo{year}{1969}), \bibinfo{pages}{318--323}.
\newblock


\bibitem[System(2018)]%
        {Display++}
\bibfield{author}{\bibinfo{person}{Cambridge~Research System}.}
  \bibinfo{year}{2018}\natexlab{}.
\newblock \bibinfo{title}{Display++ LCD Monitor}.
\newblock \bibinfo{howpublished}{\url{https://www.crsltd.com/}}.
\newblock


\bibitem[White et~al\mbox{.}(1970)]%
        {white1970eye}
\bibfield{author}{\bibinfo{person}{JH White}, \bibinfo{person}{JR Hegarty},
  {and} \bibinfo{person}{NA Beasley}.} \bibinfo{year}{1970}\natexlab{}.
\newblock \showarticletitle{Eye contact and observer bias: A research note}.
\newblock \bibinfo{journal}{\emph{British Journal of Psychology}}
  \bibinfo{volume}{61}, \bibinfo{number}{2} (\bibinfo{year}{1970}),
  \bibinfo{pages}{271}.
\newblock


\end{thebibliography}
